\documentstyle[12pt]{article}
\def\RR{{\rm I\!R}} 
\def\one{{\mathchoice {\rm 1\mskip-4mu l} {\rm 1\mskip-4mu l}
        {\rm 1\mskip-4.5mu l} {\rm 1\mskip-5mu l}}}
\def\CC{{\mathchoice {\setbox0=\hbox{$\displaystyle\rm C$}\hbox{\hbox
        to0pt{\kern0.4\wd0\vrule height0.9\ht0\hss}\box0}}
        {\setbox0=\hbox{$\textstyle\rm C$}\hbox{\hbox
        to0pt{\kern0.4\wd0\vrule height0.9\ht0\hss}\box0}}
        {\setbox0=\hbox{$\scriptstyle\rm C$}\hbox{\hbox
        to0pt{\kern0.4\wd0\vrule height0.9\ht0\hss}\box0}}
        {\setbox0=\hbox{$\scriptscriptstyle\rm C$}\hbox{\hbox
        to0pt{\kern0.4\wd0\vrule height0.9\ht0\hss}\box0}}}}
\def\ZZ{{\mathchoice {\hbox{$\sans\textstyle Z\kern-0.4em Z$}}
        {\hbox{$\sans\textstyle Z\kern-0.4em Z$}}
        {\hbox{$\sans\scriptstyle Z\kern-0.3em Z$}}
        {\hbox{$\sans\scriptscriptstyle Z\kern-0.2em Z$}}}}
\font \fivesans  = cmss10 at 5pt
\font \sevensans = cmss10 at 7pt
\font \tensans   = cmss10
\newfam\sansfam
\textfont\sansfam=\tensans
\scriptfont\sansfam=\sevensans
\scriptscriptfont\sansfam=\fivesans
\def\sans{\fam\sansfam\tensans}
\newcommand{\hf}{{\textstyle\frac{1}{2}}}
\newcommand{\ov}[1]{\overline{#1}{\kern 2pt}}
\newcommand{\slh}[1]{{#1}\kern-0.5em/}
\newcommand{\sA}{A\kern-0.5em/\kern0.2em}
\newcommand{\sZ}{Z\kern-0.6em/}
\newcommand{\sW}{W\kern-0.7em/\kern0.3em}
\newcommand{\Di}{D\kern-0.6em/\kern0.2em}
\newcommand{\fss}{\mbox{\footnotesize$\sqrt{1/3}\,\,$}}
\newcommand{\frr}{\mbox{\footnotesize$\sqrt{1/2}\,\,$}}

\newcommand{\ApZ}{ \hf g(\sA^0+\fss\sZ)}
\newcommand{\AmZ}{\hf g(\sA^0 -\fss\sZ)}

\newcommand{\lra}[1]{\mathrel{\mathop{\longrightarrow}^{\hbox{$#1$}}}}
\newcommand{\bw}{{\textstyle\bigwedge}}
\newcommand{\Wp}{{\textstyle\bigwedge^+}}
\newcommand{\Wm}{{\textstyle\bigwedge^-}}

\newcommand{\Tr}{{\rm Tr\,}}

\newcommand{\DD}{I\kern-3.5pt D}
\newcommand{\DDi}{\DD\kern-0.6em/\kern0.2em}
\newcommand{\FF}{I\kern-3.5pt F}

\begin{document}
\begin{center}
    {\LARGE\bf Superconnections and Matter}\\[1cm]
    {\large G.\ Roepstorff}\\[5mm]
    Institute for Theoretical Physics\\
   RWTH Aachen\\
    D-52062 Aachen, Germany\\
    email:\ roep@physik.rwth-aachen.de\\[2cm]
\end{center}
\begin{quote}
{\bf Abstract}. In a previous paper, the superconnection formalism 
was used to fit
the Higgs field into a $U(n)$ gauge theory with particular emphasis on
the $n=2$ case, aiming at the reconstruction of certain parts of the
Standard Model. The approach was meant as an alternative to the one that relies on
non-commutative geometry. We continue now this work by including 
matter fields (leptons) and argue in favor of a new ingredient, the right-handed 
neutrino field. It turns out that the familiar electro-weak interactions
appear naturally, provided the action 
$\int d\tau\,\overline{\Psi}\,i\DDi\Psi$
is formed with an Dirac operator $\DDi$ associated to a superconnection $D+L$
on some twisted spinor bundle. 
\end{quote}

\section{Introduction}
We resume the investigation and reconstruction of the Standard Model begun in
[1] where we concentrated on the bosonic sector. We briefly review the setup 
and the notation. The Euclidean spacetime $M$ is looked upon as some oriented 
Riemannian manifold of arbitrary dimension. For the purpose of this paper, 
however, the dimension has to be even. We let $\Gamma$ denote the algebra
of smooth complex functions on $M$. If $B$ is some complex vector bundle over 
$M$, we write $\Gamma(B)$ for the $\Gamma$-module of smooth sections 
$s:M\to B$. An operator $A$ on $\Gamma(B)$ is said to be {\em local\/} 
if it commutes with the action of $\Gamma$, i.e., $[A,f]=0$ for $f\in\Gamma$.
Typically, a differential operator is nonlocal.

Supposing that there be given a gauge connection $d_A$ on a complex Hermitian
vector bundle $V$ of rank $n$ with structure group $G\subset U(n)$, we may locally write 
$d_A=d+A$ with $A$ the connection one-form or gauge potential. The key idea is 
to pass to the induced connection $D$ on the exterior algebra $\bw V$. 
The reason is: the structure of $\bw V$ as a superbundle allows to extend 
$D$ to a superconnection $D+L$ involving the Higgs field
$L$ which is thought of as some local operator of odd type,
\begin{equation}
    L\ :\ \Gamma(\bw^\pm V)\to\Gamma(\bw^\mp V)\,, \label{L}
\end{equation}
satisfying $L=-L^*$.

In the present paper, we shall be concerned primarily with connections and
superconnections on twisted spinor bundles. We shall introduce these notions
in Sections 4 and 5 before we explain the concept of generalized Dirac
operators in Section 6. This will allow us to incorporate fermions into a
gauge theory und to study their interaction with the gauge potential $A$ and 
the Higgs field $L$ simultaneously. The reader should be warned. 
Some mathematical aspects --- mainly of interest to the geometer ---
 are left out and conditions are not
listed if they play no role in our approach. 

What really counts, in our view and in the present approach, is the natural $\ZZ_2$-grading 
(or chirality) of the spinor bundle $S$, provided the dimension of $M$ is even. We also rely 
on the fact that $\bw V$ is a superbundle which makes the twisted spinor bundle 
$E=S\otimes\bw V$ a superbundle as well. In recent years, the very notion of superbundles 
has transformed our understanding of what should be termed a connection or a
Dirac operator. Superbundles ultimately provide the right structure needed
for the concept of a superconnection. Because of its intrinsic beauty, the physical
significance should not come here as a surprise.

For the color-neutral sector of the Standard Model, the subject of this
paper, we take $G=U(2)$ as
the relevant gauge group, choose some principal $G$-bundle $P$, and construct
the associated $G$-bundle $V=P\times_G\CC^2$ with typical fiber $\CC^2$
carrying the fundamental representation of $U(2)$. Below, we shall refer to
$$
            e_1=\pmatrix{1\cr0\cr},\qquad e_2=\pmatrix{0\cr1\cr}\ .
$$
as the canonical basis in $\CC^2$. Having passed to the exterior algebra 
$$
            \bw V = P\times_G\bw\CC^2\,.
$$
we may represent the Higgs field $L$ as a $4\times4$ matrix 
\begin{equation}
  \label{hig}
  L=\pmatrix{0&i\Phi^*\cr i\Phi&0\cr}\,,\qquad 
  \Phi=\pmatrix{\Phi_1&\Phi_3\cr\Phi_2&\Phi_4\cr}
\end{equation}
with respect to the induced basis $e_I$ in $\bw\CC^2$, the index $I$ 
running over the subsets of $\{1,2\}$ including the empty set $\emptyset$.
Assume that basis 
vectors\footnote{Here, $e_\emptyset$ and $e_{12}=e_1\wedge e_2$ stand 
for the units in $\bw^0\CC^2=\CC$ and $\bw^2\CC^2\cong\CC$ respectively.} 
are arranged in the order $e_\emptyset,\ e_{12},\ e_1,\ e_2$. Then 
$$
   \pmatrix{\Phi_1\cr\Phi_2}\qquad \mbox{and}\qquad
   \pmatrix{\Phi_3\cr\Phi_4}
$$
transform as doublets under the gauge group with opposite hypercharges.

\section{The Minimal Model}

The model we propose is minimal in several ways:
\begin{itemize}
\item  Nature is assumed to provide essentially one Higgs doublet
$$
              \phi=\pmatrix{\phi_1\cr\phi_2},
$$
transforming according to the fundamental representation of $U(2)$, to which
all four components of $\Phi$ are linearly related:
\begin{equation}
  \label{mini}
    \Phi=\pmatrix{\bar{h}_1\phi_1& -h_2\bar{\phi}_2\cr
                  \bar{h}_1\phi_2& \phantom{-}h_2\bar{\phi}_1\cr}
    \qquad (h_i\in\CC).
\end{equation}
\item The Euclidean bosonic action is\footnote{In [1] we used $\mu^2C$ in
place of $\mu^2$ where $C$ was some constant gauge invariant operator.
However, since we are working with one Higgs doublet only, there is no 
loss in generality if we put $C=\one$ in the present situation.}
\begin{equation}
  \label{sb}
           S_B=g^{-2}\|\FF+\mu^2\|^2=g^{-2}\int d\tau\,|\FF+\mu^2|^2
\end{equation}
with $\FF=(D+L)^2$ the generalized curvature, $g$ the universal
gauge coupling, and $\mu>0$ setting the mass scale.
\end{itemize}
There is no need to introduce $g$ at other places than in front of the
action. However, to facilitate comparison with the Standard Model and
to remove $g$ from the kinetic terms of the action functional
we rescale the gauge potential $A$ and the Higgs field $L$, thus
giving trilinear interaction terms an extra factor $g$ and 
quartic terms a factor $g^2$. After that we absorb $g$ in
the basis $-it_a$ of the Lie algebra and in the parameters $h_i$ of the
Higgs field to preserve the relations $d_A=d+A$ and $\DD=D+L$.

Note the decomposition
$$
   |\FF+\mu^2|^2=|F|^2+|DL|^2+|L^2+\mu^2|^2
$$   
(details in [1]) where $F=D^2$ is the curvature of the gauge connection 
$D$ and $DL$ the covariant derivative of the Higgs field $L$.

The constants $h_i$ are treated as free parameters proportional to
$g$ and subject to the normalization condition
\begin{equation}
  \label{nor}
  |h_1|^2+|h_2|^2=\hf g^2
\end{equation}
which guarantees the correct behavior of the kinetic term, i.e.
$$
   g^{-2}|dL|^2= g^{-2}\Tr dL^*dL=(\partial_\mu\phi)^*(\partial^\mu\phi).
$$
The structure of $L$ has relation to the CAR algebra
\begin{equation}
  \label{car}
   \{b_i,b_k^*\}=\delta_{ik}\,,\quad \{b_i,b_k\}=\{b_i^*,b_k^*\}=0
   \qquad(i,k=1,2)
\end{equation}
formed by the operators $b_ie_I=e_i\wedge e_I$ and their adjoints. Indeed,
$$
    b_1=\left(\begin{array}{rrrr}
        0&0&0&0\\ 0&0&0&1\\ 1&0&0&0\\ 0&0&0&0\end{array}\right)\ ,\qquad
    b_2=\left(\begin{array}{rrrr}
        0&0&0&0\\ 0&0&-1&0\\ 0&0&0&0\\ 1&0&0&0\end{array}\right)
$$
and thus
\begin{equation}
  \label{lph}
     L=ih^*(\phi_1b_1+\phi_2b_2)+i(\phi_1b_1+\phi_2b_2)^*h
\end{equation}
with $h$ some constant gauge invariant operator on $\bw\CC^2$ (in fact 
the most general one):
$$
             h=\mbox{diag}\,(h_0,h_2,h_1,h_1)\,.
$$
Like any gauge invariant operator, $h$ may be regarded a function of
$$
            q=b_1b_1^*+b_2b_2^*=\mbox{diag}\,(0,2,1,1)\,.
$$
Since the value of $h_0$ does not enter the expression (\ref{lph}) for $L$,
we may take $h_0=0$.

The ansatz (\ref{sb}) gives rise to a quartic Higgs potential:
\begin{equation}
  \label{pot}
  V(\phi)=g^{-2}|L^2+\mu^2|^2=g^{-2}\Tr(L^2+\mu^2)^2
         =V_0+\frac{\lambda}{4}(\phi^*\phi-r)^2\,.
\end{equation}
Apart from the irrelevant constant $V_0$, the two other constants are
given by
\begin{equation}
  \label{con}
   \lambda g^2=8(|h_1|^4+|h_2|^4)\,,\qquad \lambda r=4\mu^2\,.
\end{equation}
Note in particular that both $\lambda$ and $r$ are positive. Moreover,
\begin{equation}
  \label{lam}
                      g^2\le\lambda\le 2g^2 
\end{equation}
in view of (\ref{nor}). The Higgs potential predicts a breakdown of symmetry,
leaving a residual $U(1)$ gauge symmetry of electromagnetism. Apart from
loop corrections the ground state is characterized by $\phi^*\phi=r$,
and choosing the unitary gauge, we are finally left with
$$
    \phi_1=2^{-1/2}\varphi + r^{1/2},\qquad \phi_2=0
$$   
where $\varphi$ is the observable neutral scalar Higgs field. By expanding
the Higgs potential to second order we extract its mass:
\begin{equation}
  \label{mass}
  V(\phi)=V_0+\hf m_H^2\varphi^2+O(\varphi^3),\qquad m_H=2\mu\,.
\end{equation}
We refer to the operator
\begin{equation}
  \label{lc}
              L_c =ir^{1/2}(h^*b_1+b_1^*h)  
\end{equation}
as the `condensate' and also write $r=v^2/2$ where $v$ is, by convention, 
the symmetry breaking parameter of the Standard Model.


\section{Masses of the Vector Bosons and the Higgs Particle}

The analysis in [1] has shown that there exists a basis $-it_a$
($a=1,\dots,4$) in the Lie algebra {\bf u(2)} adapted to the
representation on $\bw\CC^2$ in the following sense. If $T_a$ are
the generators representing $t_a$ on $\bw\CC^2$, then 
$$
            \Tr T_aT_b=\hf g^2\delta_{ab}\,.
$$
Such a basis is given by
$$
    t_a=\hf g\sigma_a\quad(a=1,2,3),
    \qquad t_4=-\hf g\sqrt{1/3}\,\one_2
$$
where the $\sigma_a$ denote the Pauli matrices.
Then the mass matrix\footnote{A matrix whose eigenvalues are the masses 
squared} $m^2$ of the four vector bosons has matrix elements
\begin{equation}
  \label{tra}
          m^2_{ab}=2g^{-2}\Tr[T_a,L_c][T_b,L_c]  \qquad(a,b=1,\ldots 4)
\end{equation}
dependent on the condensate $L_c$. When using the representation
$T_a =t_a^{ik}b_ib^*_k$
together with (\ref{lc}) and (\ref{car}) we observe an important fact:
the trace on the left of Eq.\ (\ref{tra}) is proportional to 
$$
           \Tr qhh^* = 2(|h_1|^2+|h_2|^2)
$$
and hence, by virtue of the condition (\ref{nor}), the mass matrix of 
the vector bosons does not depend on our choice of $h$. In fact, we are 
left with a simple expression:
$$
        m^2_{ab}=r\{t_a,t_b\}^{11}\,.
$$
Here $\{\cdot\}^{11}$ means that we have projected the $2\times 2$ matrix
$\{t_a,t_b\}$ onto its 11-component. Written out we get
$$
  m^2 =\hf g^2r\pmatrix{1&0&0&0\cr 0&1&0&0\cr 
       0&0&1&-\sqrt{1/3}\cr 0&0&-\sqrt{1/3}&1/3\cr}
$$
The eigenvalues of this matrix provide the masses of the four vector bosons:
$$
\begin{tabular}[t]{lll}
 mass squared of the $W^\pm$ &:& $m_W^2=rg^2/2$\\
 mass squared of the $Z^0$ &:& $m_Z^2=2rg^2/3$\\
 mass squared of the $\gamma$&:& 0
\end{tabular}
$$
The eigenvalue $rg^2/2$, when identified with $m_{W}^2$, together with 
(\ref{con})
and (\ref{mass}) leads to an expression for the Higgs coupling constant,
$$
              \lambda =\frac{g^2m_H^2}{2m_W^2}
$$
and with (\ref{lam}) to inequalities for the Higgs mass:
\begin{equation}
  \label{ineq}
                \sqrt{2}\,m_W\le m_H\le 2m_W\,.
\end{equation}
The eigenvectors corresponding to the eigenvalues $2rg^2/3$ and 0 are written
in terms of the Weinberg angle $\theta_W$:
\begin{eqnarray*}
t_z&=&\cos\theta_W\,t_3-\sin\theta_W\,t_4,\qquad\cos\theta_W=\hf\sqrt{3}\\[2mm]
t_0&=&\sin\theta_W\,t_3+\cos\theta_W\,t_4,\qquad\sin\theta_W=\hf\,.
\end{eqnarray*}
The change of basis leads to the photon field $A^0_\mu(x)$ and the field
$Z_\mu(x)$ of the $Z$ particle such that $Zt_z+A^0t_4=A^3t_3+A^4t_4$. Hence,
\begin{eqnarray*}
  Z &=& \cos\theta_W\, A^3 -\sin\theta_W\,A^4\\ 
 A^0&=& \sin\theta_W\, A^3 +\cos\theta_W\,A^4\,.
\end{eqnarray*}
It is gratifying to see that these results coincide with our analysis in [1],
except for the lack of a definite prediction of the Higgs mass, though the 
value $m_H=2m_W$, which previous reasoning has led us to, is within reach.
It marks the upper limit of (\ref{lam}) obtained if either $h_1=0$
or $h_2=0$. As we shall see, $h_1=0$ corresponds to the assertion that the
Higgs mechanism gives zero mass to the neutrino. Even if it turns out that the
neutrino has non-zero mass, the relation $m_H=2m_W$ remains valid to
great accuracy.


\section{Spinor Bundles and Their Duals}

Let $M$ be an oriented Riemannian manifold of dimension $2m$ and $C(M)$ be its 
Clifford bundle. Recall that the bundle $C(M)$ has the complexified Clifford algebra 
$C(T^*_xM)\otimes\CC$ as fibre at $x\in M$ and that the real Clifford algebra 
$C(V)$ associated to any real Hermitian vector space $V$  is characterized (up to 
isomorphisms) by the following universal property: any linear map $c:V\to A$ satisfying 
$c(v)^2+(v,v)=0$, where $A$ is an associative algebra with unit,  can be extended 
to an algebraic homomorphism $c:C(V)\to A$. Note that the Clifford algebra $C(M)$ is
$\ZZ_2$-graded with grading automorphism $v\mapsto -v$. Details may
be found in Refs.\ [2] and [3].                 

A {\em left Clifford module\/} on $M$ is a $\ZZ_2$-graded complex vector bundle $S$
with a left action of $C(M)$ respecting the grading, i.e.,
$$
      C^+(M)\cdot S^\pm\subset S^\pm,\qquad  C^-(M)\cdot S^\pm\subset S^\mp
$$
A {\em right Clifford module\/} on $M$ is defined in a similar manner. To any left Clifford
module $S$ we associate a right Clifford module $\overline{S}$, called the {\em dual\/}
 of $S$, in a natural way:
\begin{enumerate}
\item The vector space ${\overline S}_x$ at $x\in M$ is the dual of $S_x$.
\item The right action of the Clifford algebra $C(M)$ is defined in such a way that the
          canonical linear function 
          $
          \overline{S}\otimes S\to M\times\CC,\quad\overline{s}\otimes s\mapsto\overline{s}s
          $
          factorizes as follows:
          \begin{equation}
          \overline{S}\otimes S\to \overline{S}\otimes_{C(M)} S\to M\times\CC       \label{fac} 
          \end{equation}
\item The grading is reversed: $\overline{S}{\kern 2pt}_x^\pm$ is the dual of $S_x^\mp$.
\end{enumerate}
We shall be dealing mainly with sections of bundles. Our primary concern is with the algebra
$$
       {\cal C}=\Gamma(C(M))
$$
of sections of the Clifford algebra. It should be clear then that
$\Gamma(S)$ is a left ${\cal C}$-module while $\Gamma(\overline{S})$ is a right
${\cal C}$-module. By construction, $\overline{\psi}\psi\in\Gamma$ for $\psi\in\Gamma(S)$
and $\overline{\psi}\in\Gamma(\overline{S})$. The condition (\ref{fac}) says that
\begin{equation}
         (\overline{\psi}\cdot a)\psi = \overline{\psi}(a\cdot\psi) \qquad a\in{\cal C}
        \label{com}
\end{equation}
which defines $\overline{\psi}\cdot a\in\Gamma(\overline{S})$ uniquely.

Let $c: T^*M \to {\rm End\,}S$ be a spin$^c$-structure on $M$. In detail: $S$ is
some complex vector bundle of rank $2^m$, called the
{\em spinor bundle\/}, and the bundle map $c$ satisfies
\begin{equation}
                                     c(v)^2+(v,v)=0\qquad v\in T^*M\ . \label{c}
\end{equation}
The significance of a spin$^c$-structure is that the map $c$ extends to an irreducible Clifford
action on $S$, i.e., by the above universal property, there is a unique algebraic homomorphism 
$c: C(M) \to {\rm End\,}S$ (in fact an isomorphism) compatible with (\ref{c}). In other words, $S$ is a left
Clifford module with the action of $C(M)$ given by $a\cdot s=c(a)s$. We will
show that $S$ is $\ZZ_2$-graded and that the action of $C(M)$ respects the grading.

With $dx^k$ some oriented orthonormal frame in $T_x^*M$ and 
$\gamma^k=c(dx^k)$ we obtain the relations
$$
   \{\gamma^j,\gamma^k\}=-2\delta^{jk},\qquad j,k=1,\ldots,2m\ .
$$
Generally speaking, the global operators $\gamma^k$ depend on $x\in M$ and may only
locally be represented by constant $\gamma$-matrices.

The notion of a spin$^c$-structure is the complex analogue of the notion of a (real)
spin structure. It seems to be the important concept for our purpose.
Not every manifold admits a spin$^c$-structure. The conditions for a manifold to be
spin$^c$ may be stated in terms of cohomology: (1) the first Stiefel-Whitney class
vanishes (which makes $M$ orientable) and (2) the second Stiefel-Whitney class is the
mod 2 reduction of an integral class [2].

Consider the {\em chirality\/} operator $\gamma_5  =  i^{m}\gamma^1\gamma^2
\cdots\gamma^{2m}$ whose construction is independent of the frame. It satisfies 
\begin{equation}
    \gamma^2_5=1,\quad \{\gamma_5,\gamma^k\}=0\ .
\end{equation}
A $\ZZ_2$-grading on the spinor bundle $S$ is defined by 
$$
                   S^\pm =\{s\in S\,|\, \gamma_5s=\pm s\}\ .
$$
It is clear that the subbundles $S^\pm$ have the same rank.
Since $\gamma_5$ anticommutes with $\gamma^k$, the (left) action of $C(M)$ on $S$
respects the grading. It goes without saying that the dual spinor bundle $\overline{S}$
is $\ZZ_2$-graded with a graded (right) action of $C(M)$.

Sections $\psi\in\Gamma(S)$ are referred to as {\em Dirac fields}. We call
$\psi\in\Gamma(S^\pm)$ right-handed ($+$ sign) resp.\ left-handed ($-$ sign) 
and suggestively write $\psi=\psi_R+\psi_L$ for the decomposition of
$\psi\in\Gamma(S)$ into Dirac fields of definite chirality. Analogously, we write
$\overline{\psi}=\overline{\psi}_R+\overline{\psi}_L$ for $\overline{\psi}\in\Gamma(\overline{S})$.
Since $S$ and $\overline{S}$ carry opposite gradings,
$$
      \overline{\psi}\psi=\overline{\psi}_R\psi_L+\overline{\psi}_L\psi_R\in\Gamma\ .
$$
When it comes to application in quantum field theory, we should always be aware of an
important difference between the Euclidean and the Minkowski approach.
In the Euclidean formulation, which is the point of view adopted here, 
$\psi$ and $\overline{\psi}$ are by no means related. Quite on the contrary, 
since they belong to different spaces, 
$\psi$ and $\overline{\psi}$ are independent variables. On the other hand,
$\psi$ and $\overline{\psi}$ {\em are\/} connected in the Minkowski version of operator 
field theory. 

It is common practice to write $\slh{\partial}=\gamma^\mu \partial_\mu$ in local 
coordinates. If $d\tau$ is the volume element of the manifold $M$, the Euclidean action $S_E$
of a free Dirac particle of mass $m$ reads:
\begin{equation}
          S_E =\int_M d\tau \, \overline{\psi}(i\slh{\partial}-m)\psi
         \label{dirac}
\end{equation}
Observe that $(\slh{\partial}\psi)_R = \slh{\partial}\psi_L$ and $(\slh{\partial}\psi)_L
 = \slh{\partial}\psi_R$ and thus
$$
  \overline{\psi}(i\slh{\partial}-m)\psi   =\overline{\psi}_Ri\slh{\partial}\psi_R 
                                          +\overline{\psi}_Li\slh{\partial}\psi_L
                                          -m(\overline{\psi}_R\psi_L+\overline{\psi}_L\psi_R)
$$
If $m=0$, the action $S_E$ is invariant under global $U(1)\times U(1)$ transformations. 
This symmetry referred to as {\em chiral invariance} is broken down to $U(1)$
by the mass term.

We may look upon $\slh{\partial}$ as the simplest example of a Dirac operator. It has
two crucial properties: (1) it is an odd operator, i.e., it maps $\Gamma(S^\pm)$ into 
$\Gamma(S^\mp)$, and
(2) satisfies $[\slh{\partial},f]=c(df)$ for $f\in\Gamma$. These properties serve as the 
point of departure for the generalized notion of a Dirac operator. The extended concept
will be very fruitful when gauge couplings to $A$ and Yukawa couplings to $L$ are taken into
account.

\section{Twisted Spinor Bundles}

To incorporate the graded vector bundle $\bw V$ of a $U(n)$ Yang-Mills-Higgs
system means giving further degrees of freedom to the spinor field.
The idea is then to work with the {\em twisted spinor bundle\/} 
$$                      E=S\otimes\bw V   $$ 
with Clifford action $c\otimes 1$. No confusion can arise if we continue to write $c$ in
place of $c\otimes 1$. The induced grading of the bundle $E$ is described 
as follows:
\begin{eqnarray*}
              E_+ &=& E^+_R\oplus E^-_L\ ,\qquad E_R^\pm :=S^+\otimes\bw^\pm V\\
              E_-  &=& E^+_L\oplus E^-_R\ , \qquad E_L^\pm :=S^-\otimes\bw^\pm V\ .
\end{eqnarray*}
This makes $E$ a left Clifford module. Lateron, calculations will be facilitated by decomposing
the spinor field $\Psi\in\Gamma(E)$ in the following manner: 
$$
     \Psi  =\Psi_++\Psi_-\ , \qquad 
     \Psi_+=\pmatrix{\psi^+_R\cr\psi^-_L}\in \Gamma(E_+)\ ,\qquad
     \Psi_-  =\pmatrix{\psi^+_L\cr\psi^-_R} \in \Gamma( E_-) \ .                                                  
$$
This will then suggest a matrix representation of Dirac operators. Observe that such
decompositions of fields and operators always refers to the $\ZZ_2$-grading of $\bw V$ 
rather than to the grading of $E$. 

The dual $\overline{E}$ as a right Clifford module is now well defined. To see its detailed 
structure we proceed in steps. Firstly, since $\bw V$ is regarded as a trivial left Clifford 
module,  $\ov{\bw V}$ is a trivial right Clifford module with grading
$$
                      \ov{\bw V}^\pm =\bw^\mp V^*
$$
where $V^*$ is the dual of the (ungraded) vector space $V$. Secondly, from
$$
      \ov{E}= \overline{S\otimes\bw V}=\overline{S}\otimes\overline{\bw V}
$$
one learns that
\begin{eqnarray*}
               \overline{E}_+ &=&  \ov{E}^+_L\oplus  \ov{E}^-_R\ ,
           \qquad  \ov{E}_R^\pm :=\ov{S}^+\otimes\bw^\pm V^*\\
               \overline{E}_-  &=&  \ov{E}^+_R\oplus  \ov{E}^-_L\ , 
           \qquad  \ov{E}_L^\pm :=\ov{S}^-\otimes\bw^\pm V^*\ .
\end{eqnarray*} 
It will prove convenient to decompose the dual spinor field $\overline{\Psi}\in\Gamma(\overline{E})$ as follows:
$$
\overline{\Psi}=\overline{\Psi}_++\overline{\Psi}_-\ ,\qquad
 \overline{\Psi}_+=(\ov{\psi}^+_L,\ov{\psi}_R^-)\ ,\qquad
 \overline{\Psi}_-=(\ov{\psi}^+_R,\ov{\psi}_L^-)\ .
$$
Suppose we want to fit the electron-neutrino system into the present framework by
defining a suitable Clifford bundle $E$. As is appropriate for the electro-weak theory, 
we choose some rank 2 vector bundle $V$ with structure group $U(2)$ as in
Section 1. The rank of the graded vector bundle $\bw V$ is four, 
signalizing that there are four basic Dirac fields. In standard notation, these fields are 
$e_R$, $e_L$, $\nu_{eL}$, and $\nu_{eR}$. We combine them to one spinor field 
$\Psi\in\Gamma(E_+)$. Restriction to $E_+$ here means that we identify the
parities in $\bw V$ and $S$, i.e.,
\begin{eqnarray*}
   \psi^+_R &=&\pmatrix{\nu_{eR}\cr e_R}\in\Gamma(E_R^+)\ ,\qquad
   E^+_R=S^+\otimes\bw^+ V\ ,\qquad \bw^+ =\bw^0\oplus\bw^2 \\[3mm]
 \psi^-_L &=&\pmatrix{\nu_{eL}\cr e_L}\in\Gamma(E_L^-)\ ,\qquad
   E^-_L=S^-\otimes\bw^- V\ ,\qquad \bw^- =\bw^1 \ .
\end{eqnarray*}
Investigating the behavior under transformations $u\in U(2)$, we find: 
\begin{enumerate}
\item The right-handed neutrino field $\nu_{eR}$ transforms trivially because $\bw^0u=1$.
           In particular, its hypercharge is zero.
\item The right-handed electron field $e_R$ transforms as an $SU(2)$ singlet, but has
          hypercharge $-2$ as follows from $\bw^2u={\rm det\,}u$.
\item The pair consisting of the left-handed neutrino field $\nu_{eL}$ and the
          left-handed electron field $e_L$ transforms as a $SU(2)$-doublet and has
          hypercharge $-1$. This is a consequence of $\bw^1u=u$.
\end{enumerate}
Dually, if $\overline{\Psi}=\in\Gamma(\overline{E}_-)$, we have 
that $\ov{\psi}^+_L=\ov{\psi}^-_R=0$ and
\begin{eqnarray*}
\ov{\psi}^+_R &=&(\overline{\nu}_{eR},\overline{e}_R)\in\Gamma(\ov{E}^+_R)\ ,
                                     \qquad \ov{E}^+_R=\ov{S}^+\otimes\Wp V^*\\
\ov{\psi}^-_L &=&(\overline{\nu}_{eL},\overline{e}_L)\in\Gamma(\ov{E}^-_L)\ ,
                                     \qquad \ov{E}^-_L=\ov{S}^-\otimes\Wm V^*
\end{eqnarray*}
Why consider a right-handed neutrino field $\nu_{eR}$ and its dual $\overline{\nu}_{eR}$, 
allegedly unobserved in nature? There is an obvious answer to this question.

Both the electron and the neutrino receive their masses by the Higgs
mechanism. No doubt, a fermionic action without mass terms and no 
right-handed neutrino field is consistent. One simply assumes that
$\nu_{eR}=0$ from the beginning. If however $m_{\nu_e}\ne 0$ after
sponaneous breaking of the $U(2)$ symmetry, the neutrino receives a new degree
of freedom: $\nu_{eR}\ne 0$. This additional spinorial degree of freedom cannot be
created out of nothing. It can certainly not be provided by the Higgs degrees
of freedom. It therefore seems reasonable to incorporate the right-handed
neutrino field from the beginning.

Recent experimentally observed neutrino oscillations indicate that neutrinos
may indeed be massive. To summarize our point of view, the neutrino is in no 
way different from the electron.


\section{Dirac Operators}

On a Riemannian manifold $M$, there is precisely one torsionfree and metric connection
$\nabla$ on the tangent bundle $TM$, called the {\em Levi-Civita connection\/}:
$$
              \nabla \ : \ \Gamma( TM)\to \Gamma(T^*M\otimes TM)\ .
$$
In local coordinates, $\nabla$ may be written in terms of the Christoffel symbols 
$\Gamma^k_{ij}$:
$$
   \nabla= dx^i\otimes\nabla_i\ ,\qquad 
   \nabla_i\partial_j=\Gamma^k_{ij}\partial_k
$$
The Levi-Civita connection induces connections on a variety of bundles which are also
referred to as Levi-Civita connections. In particular, there are canonical connections
on the Clifford bundle $C(M)$ and the spinor bundle $S$ which we also denote by
$\nabla$ for simplicity. Since the spinor bundle is a ${\cal C}$-module, the covariant 
derivative
\begin{equation}
        \nabla\ :\ \Gamma(S)\to\Gamma(T^*M\otimes S)
\end{equation}
is defined so as to be compatible with the Levi-Civita connection of the Clifford bundle:
\begin{equation}
        [\nabla,c(a)]=c(\nabla a) \ ,\qquad a\in{\cal C} \ .                 \label{LC}
\end{equation}
The latter equation says that the Levi-Civita connection on $S$ is a {\em Clifford 
connection\/} (which is a more general concept). 
Note that, in Eq.\ (\ref{LC}), $\nabla a$ is viewed as a $\cal C$-valued one-form.
Again, the Levi-Civita connection on $S$ may locally be written in terms of the
Christoffel symbols:
$$
       \nabla = dx^i\otimes\nabla_i\ ,\qquad
       \nabla_i =\partial_i+{\textstyle\frac{1}{4}}
                        \Gamma^k_{ij}\gamma^j\gamma_k\ ,\qquad \gamma^i=c(dx^i)\ .
$$           
This represention would also hold if we had started from an affine connection on $TM$.

Given a connection $D$ on the superbundle $\bw V$, there is a natural 
extension\footnote{While products like $\nabla\otimes 1$ and $1\otimes D$
separately are meaningless, their sum has a meaning owing to the consistency
condition $[\nabla,f]=[D,f]$ ($f\in\Gamma$).} 
to a connection $D=\nabla\otimes 1+1\otimes D$ on the twisted spinor bundle 
$E=S\otimes \bw V$. We then define the associated Dirac operator $\Di$ by means of
the following composition of maps,
$$
     \Di\ :\ \Gamma(E)\lra{D}\Gamma(T^*M\otimes E)\lra{c}\Gamma(E)\ ,
$$
so that, in local coordinates,
$$
           D=dx^i \otimes D_i\quad\Rightarrow\quad \Di = \gamma^i D_i   \ .                       
$$
The operator $\Di$ on $\Gamma(E)$ thus obtained is justly called a Dirac operator. 
The reason is that it satisfies two fundamental conditions:
\begin{enumerate}
\item $\Di$ is an odd operator, i.e., it maps $\Gamma(E_\pm)$ into 
          $\Gamma(E_\mp)$.
\item $\Di$ respects the $\Gamma$-module structure of $\Gamma(E)$, i.e.,
          $[\Di,f]=c(df)$ for all $f\in\Gamma$.
 \end{enumerate}
Dirac operators are at the heart of modern Riemannian geometry [2]. They establish a close
relationship between topology and curvature and provide the basis for index theorems of
elliptic operators. Though Dirac was first to construct an example, 
the general concept is due to Atiyah and Singer [4] who also realized 
the vital importance of this notion in geometry.

The Dirac operator defined above does not yet include the Higgs field. However,
the passage from $D$ to $\Di$ can be extended\footnote{In local coordinates,
it means replacing $dx^i$ by $\gamma^i$ everywhere in $\DD$.} to 
superconnections $\DD=D+L$. If $L$ is scalar (a 0-form), the construction 
of the associated Dirac operator is particularly simple:
$$
          \DD=D+L\qquad\Rightarrow \qquad 
          \DDi=\Di+L=\pmatrix{\Di^+& i\Phi^*\cr i\Phi&\Di^-\cr}\,.
$$
It still obeys 
$$
        \DDi\colon \Gamma(E_\pm)\to\Gamma(E_\mp)\,.
$$
What enters the fermionic part of the Euclidean action is but the 
restriction $\DDi\colon \Gamma(E_+)\to\Gamma(E_-)$:
\begin{equation}
          S_F=\int_M d\tau\, \overline{\Psi}\,i\DDi\Psi\ ,\qquad
                     \Psi\in\Gamma(E_+)\ .   \label{SE}
\end{equation}
Writing out the integrand we get
\begin{equation}
  \ov{\Psi}\,i\DDi\Psi =  \ov{\psi}^+_R \,i\Di^+\psi^+_R
                         +\ov{\psi}^-_L\,i\Di^-\psi^-_L       
                         -\ov{\psi}^-_L \,\Phi\psi^+_R 
                         -\ov{\psi}^+_R \,\Phi^*\psi^-_L      \label{lag}
\end{equation}
While the first two terms describe the gauge part including effects of 
the Levi-Civita connection, the last two terms represent the Yukawa 
interactions. It becomes immediately obvious that the parameters $h_1$
and $h_2$ are the two relevant Yukawa coupling constants of the theory and
that the condensate $L_c$ gives rise to a mass matrix of the fermions
(diagonal in the present situation).

Quantization of a chiral $U(2)$ gauge theory, however, is beset of difficulties
the origin of which is the well-known {\em local anomaly}. Therefore, it seems
unreasonable to propose the model as it stands as the basis of a consistent 
quantum field theory. What we have at our hand most probably is some kind of 
effective
field theory, valid perhaps only on tree level. The deficit calls for an
extension to a larger gauge group and an inclusion of further degrees
of freedom (quarks). It is generally accepted that
without quarks there will be no anomaly-free formulation.


\section{The $U(2)$ Gauge Theory of Electrons \\ and Neutrinos}

To test the relevance of the superconnection concept for the Standard Model 
[5,6] 
we shall disregard effects coming from the Levi-Civita connection and 
take the flat Euclidean space $\RR^4$ as the underlying manifold $M$. 
Since the principal $G$ bundle $P$ then admits global sections, the bundle is 
trivial, $P=M\times G$, and so are all associated vector bundles like $V$ and
$\bw V$.
Intricacies of global topology are absent.

With the gauge group $U(2)$ broken down to the residual $U(1)$ group of
electromagnetism the connection 1-form on $V$ changes to the
following $2\times 2$ matrix
$$
 A =ig\pmatrix{\fss Z  &\frr W^+\cr 
               \frr W^-&\hf(A^0-\fss Z)\cr}
$$
involving three massive vector fields in addition to the massless photon field
$A^0$. On the exterior algebra $\bw V$ we have the implied $4\times4$
representation
$$
  \hat{A}=\pmatrix{\hat{A}^+& 0\cr 0&\hat{A}^-\cr}\,,\qquad
  \hat{A}^+=\pmatrix{0&0\cr 0& \mbox{tr}\,A\cr}\,,\qquad
  \hat{A}^-=A\,.
$$
Passage to the unitary gauge means that the Higgs degrees of freedom
are reduced to a residual neutral scalar field $\varphi$:
$$
    L=i\pmatrix{0&\Phi^*\cr \Phi&0\cr}\ ,\qquad
   \Phi =\frr\pmatrix{\bar{h}_1(\varphi+v)& 0\cr 
                   0&h_2(\varphi+v)\cr}\,.
$$
The Dirac operator associated with the superconnection $\DD$ is
\begin{equation}
     \DDi=\slh{\partial}+\omega\ ,\qquad
      \omega =\hat{A}+L\ .  \label{om}
\end{equation}
where $\omega$ may be written as a $4\times4$ matrix of operators acting
on the spinor space $S$:
$$
  \omega =i\left(\begin{array}{cccc}
   0 & 0    &  h_1\frr(\varphi+v) &  0                             \\[5pt]
   0 & \ApZ & 0                        & \bar{h}_2\frr(\varphi+v) \\[5pt]
   \bar{h}_1\frr(\varphi+v)& 0   &  g\fss\sZ  & g\frr\sW^+        \\[5pt]
   0 & h_2\frr(\varphi+v)  & g\frr\sW^-       &       \AmZ     
   \end{array}\right)
$$
Similarly, $\Psi$ and $\overline{\Psi}$ enter $\overline{\Psi}i\DDi\Psi$ as:
$$
    \Psi =\pmatrix{\nu_{eR}\cr e_R\cr \nu_{eL}\cr e_L\cr}\ ,\qquad
    \overline{\Psi}=(\bar{\nu}_{eR},\bar{e}_R,\bar{\nu}_{eL},\bar{e}_L)\ .
$$
The expression $\overline{\Psi}\,i\omega\Psi$ incorporates all known 
interactions of the electron-neutrino system plus the appropriate mass term:
$$
\begin{tabular}{ll}
$-\hf g(\bar{e}_R\sA^0e_R+\bar{e}_L\sA^0e_L) \qquad $&
                  electromagnetic interaction\\[5pt]
$-g\frr(\bar{\nu}_{eL}\sW^-e_L+\bar{e}_L\sW^+\nu_{eL})\qquad$&
                  weak charged current interaction\\[5pt]
$-\hf g\fss(\bar{e}_R\sZ e_R-\bar{e}_L\sZ e_L
 +2\bar{\nu}_{eL}\sZ \nu_{eL})\qquad$&
                  weak neutral current interaction\\[5pt]
$-\frr(h_2\bar{e}_L e_R+\bar{h}_2\bar{e}_R e_L)\varphi\qquad$&
                  Yukawa interaction of the electron\\[5pt]
$-\frr(h_2\bar{e}_L e_R+\bar{h}_2\bar{e}_R e_L)v\qquad$&
                  electron mass term
\end{tabular}
$$
But it also incorporates two additional terms, relevant only if the
neutrino acquires a mass, i.e., if $h_1\ne0$:
$$
\begin{tabular}{ll}
$-\frr(\bar{h}_1\bar{\nu}_{eL}\nu_{eR}
       +h_1\bar{\nu}_{eR}\nu_{eL})\varphi\qquad$&
                  Yukawa interaction of the neutrino\\[5pt]
$-\frr(\bar{h}_1\bar{\nu}_{eL}\nu_{eR}+h_1\bar{\nu}_{eR}\nu_{eL})v\qquad$&
                  neutrino mass term
\end{tabular}
$$
Let us summarize the positive results concerning the (bare) parameters of the 
model:
\begin{enumerate}
\item The Weinberg angle is given by $\sin^2\theta_W=0.25$ while the
      measured value is $\sin^2\theta_W=0.23$. This agrees rather well with
      the observed ratio $m_W/m_Z$.
\item The gauge coupling constant $g$ and the fine structure constant $\alpha$
      at the energy scale $\mu$ appear to be related by $(g/2)^2=4\pi\alpha$.
      Accepting $\alpha(\mu)^{-1}=128$ leads to the prediction $g(\mu)=0.62$,
      while $g(0)=0.65$ from the muon decay and the equation
      $\sqrt{2}\, Gm_W^2=g^2/4$ relating $g$ to the Fermi constant $G$.
\item The Higgs mass is predicted to be $m_H\approx 2m_W$.
\item The Higgs (self)coupling constant $\lambda$ is positive and 
      approximately $2g^2$.
\item The symmetry breaking parameter 
      $         v=\sqrt{2r}=2m_W/g\approx 258\ \mbox{GeV} $
      is rather close to the generally accepted value 246 GeV.
\end{enumerate}
On the negative side we mention: the fermion masses as given by the equations
$$
      m_{e}^2= r|h_1|^2\,,\qquad  m_{\nu_e}^2= r|h_2|^2
$$
render the absurd relation
\begin{equation}
  \label{abs}
     m_{e}^2+m_{\nu_e}^2=r(|h_1|^2+|h_2|^2)=\hf rg^2=m_W^2\,.  
\end{equation}
We figure there must be something fundamentally wrong with the assumption
that there exist an electroweak theory of leptons alone. Conclusion:
quarks and leptons ought to be regarded as entities of the same kind
described by a common field $\Psi$, transforming according to some 
(possibly reducible) representation of a larger gauge group.
Outlook: we shall in a forthcoming paper present a formulation 
describing all fermions of three generations. 
The relation (\ref{abs}) will be generalized so as to include all fermion
masses since the right hand side involves the trace of $M^*M$ with $M$ the mass
matrix of the fermions. Relations connecting fermion and boson masses are
typical ingredients of a superconnection model.

\vspace{1cm}
\leftline{\Large \bf References}\vspace{2mm}
\begin{enumerate}
\item G.\ Roepstorff, {\em Superconnections and the Higgs Field\/}, 
          hep-th/9801040
\item H.B.\ Lawson and M.-L.\ Michelsohn: {\em Spin Geometry\/}, Princeton University
          Press, Princeton, N.J. 1989
\item N.\ Berline, E.\ Getzler, M.\ Vergne: {\em Heat Kernels and Dirac Operators\/},
          Springer, Berlin 1996
\item M.F.\ Atiyah and I.M.\ Singer, Ann.Math.\ {\bf 87}, 484 (1968); {\bf 87}, 546 (1968);
          {\bf 93}, 119 (1971); {\bf 93}, 139 (1971)
\item S.\ Weinberg: {\em The Quantum Theory of Fields, Vol.II: Modern Applications\/}.
         Cambridge University Press 1996
\item F.\ Scheck: {\em Electroweak and Strong Interactions}, 2nd Ed.,
       Springer 1996
\end{enumerate}
\end{document}